\documentclass[conference]{IEEEtran}

\IEEEoverridecommandlockouts
\usepackage[pdftex]{graphicx}

\usepackage{mathtools,  cuted}
\usepackage{lipsum,  color}

\usepackage{mathtools}
\usepackage{graphics}

\usepackage{amsmath,  amsfonts}
\usepackage{nopageno}
\usepackage{algpseudocode}
\usepackage{float}
\usepackage{verbatim}
\usepackage{cite}
\usepackage{array}
\usepackage{amsmath}
\usepackage{icomma}

\begin{document}

\title{A New Millimeter Wave MIMO System for 5G Networks}

%
%


\author{
    \IEEEauthorblockN{Mojtaba Ahmadi Almasi\IEEEauthorrefmark{1},  Roohollah Amiri\IEEEauthorrefmark{1},  Hani Mehrpouyan\IEEEauthorrefmark{1}}\\
    \vspace{-0.5cm}
    \IEEEauthorblockA{\IEEEauthorrefmark{1}{\small {Department of Electrical and Computer Engineering,  Boise State University}}
   \{mojtabaahmadialm, roohollahamiri,\\ hanimehrpouyan\}@boisestate.edu}
   \vspace{-40pt}
   \\
   \thanks{This project is supported in part by the NSF ERAS grant award number 1642865.} 
}

\maketitle

\begin{abstract}
Millimeter Wave (mmWave) band provides a large spectrum to meet the high-demand capacity by the 5th generation (5G) wireless networks. However,  to fully exploit the available spectrum,  obstacles such as high path loss,  channel sparsity and hardware complexity should be overcome. To this end,  the present paper aims to design a new multiple-input multiple-output (MIMO) system by using lens-based multi-beam reconfigurable antennas. The proposed MIMO system uses complete lens at the transmitter and incomplete lens at the receiver. To reduce hardware complexity,  we utilize an optimal beam selection technique. Our analysis demonstrates that the proposed MIMO system along with the optimal beam selection technique increases the average signal-to-noise ratio (SNR). Also, simulations show that the system achieves full-diversity gain. 
\end{abstract}


%

\section{Introduction}
Millimeter Wave (mmWave) communications operating in $30-300$ GHz range has been proposed as one of the feasible solutions for the fifth-generation (5G) wireless networks~\cite{r1}. However,  significant path loss,  blockage,  and hardware limitations are major obstacles for the deployment of mmWave systems. To address these obstacles,  several mmWave systems have been proposed to date~\cite{r5, el2014spatially, r9, r19, almasi2018new}.

An analog beamforming mmWave system consists of one radio frequency (RF) chain,  an antenna array,  and phase shifters that connects the RF element to the antenna array~\cite{r5}. Although the system has low hardware complexity and is cost-efficient,  it is able to support only one data stream. In order to transmit multiple streams and keep the number of RF chains small,  the authors in~\cite{el2014spatially} designed hybrid beamforming system. Since the channels in mmWave frequencies are sparse\footnote{A sparse channel is defined as a channel in which the number of paths with strong gain is far less than all existing paths.},  the hybrid system  exploits the sparsity property of these channels. Nevertheless,  hybrid beamforming uses a large number of phase shifters that brings hardware complexity. In order to completely eliminate phase shifters,  the concept of beamspace multi-input multi-output (MIMO) is introduced in~\cite{r9}. Compared to the hybrid beamforming,  beamspace MIMO is affordable and exploits the sparsity of mmWave channels appropriately such that its capacity approaches those of fully-digital system. 

Both hybrid beamforming and beamspace MIMO systems successfully reduce the number of RF chains which are power-hungry and expensive elements. Although these systems are suitable for outdoor communication in mmWave frequencies,  there are three major problems with these systems~\cite{r19}. First,  they fail to overcome sparsity and low-rank channels. Second,  to achieve multiplexing gain the number of RF chains at the transmitter and the receiver should be equal. This may not be feasible in practical point-to-point communications. Third, due to the spacial structure, the hybrid beamforming and beamspace MIMO do not provide diversity gain and focus on achieving multiplexing gain. Consequently,  they cannot reduce bit error rate (BER).  To address these issue while preserving the complexity and cost of the antenna, a lens-based multi-beam reconfigurable antenna-MIMO (RA-MIMO) system\footnote{The original lens-based multi-beam antenna was designed and fabricated in~\cite{schoenlinner2002wide} for only one tapered slot antenna (TSA) array.} has been recently proposed in~\cite{r19} and~\cite{almasi2018new}. In particular,  the RA-MIMO can provide a large number of paths in mmWave band~\cite{r19, almasi2018new}. Further,  it attains full-diversity gain as well as multiplexing gain. However, the complete spherical lens, which is used in the reconfigurable antenna, requires large space at both base station and mobile user device in practice.  

In this paper,  motivated by the reconfigurable antenna in~\cite{r19},  we design a new RA-MIMO structure for mmWave systems. The new RA-MIMO deploys lower number of lenses than the counterpart in~\cite{almasi2018new} at the base station and only an incomplete spherical lens at user device. Hence,  the new RA-MIMO reduces the needed room inside mobile devices. We also utilize an optimal beam selection technique integrated with Alamouti space-time block code (STBC) to obtain full-diversity gain. The analytical finding indicates that the beam selection technique gives higher gain in comparison with random beam selection. Also, the simulation results show that the proposed RA-MIMO  is able to achieve full-diversity gain.    

The paper is organized as follows: Section~II presents the system model and the proposed MIMO. In Section~III,  an optimal algorithm is presented for beam selection. In Section~IV,  we present simulations investigating the performance of the proposed MIMO in terms of beam selection gain and BER. Section V concludes the paper.

\textbf{Notations:} Hereafter,  small letters,  bold letters and bold capital letters will designate scalars,  vectors,  and matrices,  respectively. Also,  $(\cdot)^T$ and $|\cdot|$ denote transpose operation and the absolute value of $(\cdot)$, respectively. $\mathbb{E}[\cdot]$ denotes the expected value of $(\cdot)$

\section{System Model and The Proposed MIMO}\label{sec:system}
In this section, first, the system model is described. Then, the new RA-MIMO system is proposed. Also, for the proposed RA-MIMO with random beam selection, the instantaneous SNR is calculated. 
\begin{figure}
\vspace{-0.5cm}
\hspace*{-0.65cm}
    \centering
    \includegraphics[width=0.9\columnwidth]{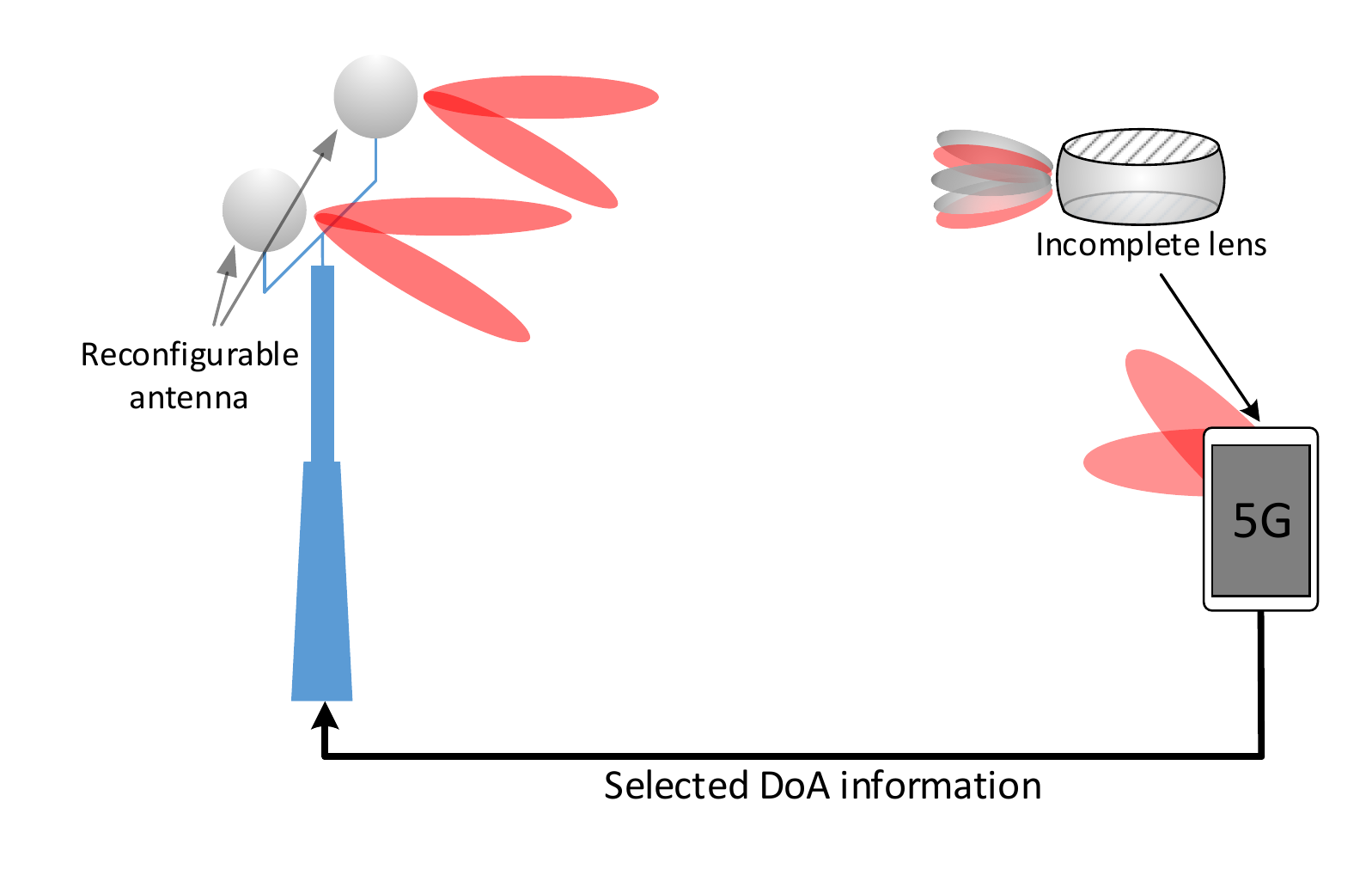}
    \vspace{-0.5cm}
    \caption{Downlink of a 5G mmWave system implemented by the proposed RA-MIMO.}
    \label{fig1}
\end{figure}
\subsection{System Model}\label{subsec:systemmodel}
The downlink of a 5G mmWave transmission between a base station and mobile user is depicted in Fig.~\ref{fig1}. The base station is equipped with $M$ reconfigurable antennas. Each reconfigurable antenna consists of four main components,  i) RF transceivers,  ii) beam selection network (BSN),  iii) multiple taper slot antenna (TSA) arrays,  and iv) spherical lens located in front of the TSAs~(see Fig.~3 in \cite{r19}.). At the user end,  it is assumed that only one reconfigurable antenna is deployed. The main reason for such assumption is to resolve the space issue in the user device. 

We consider complete spherical lens for the transmit reconfigurable antenna. According to~\cite{r19},  a complete lens is of diameter 65 mm. Therefore,  even a single lens occupies large room in the user device. To further reduce the required space at the mobile device,  we use an incomplete lens instead of the complete one according to Fig.~\ref{fig1}. 

\subsection{The Proposed Reconfigurable Antenna-MIMO}\label{subsec:proposed}

\begin{figure*}
\vspace{-0.5cm}
    \centering
    \includegraphics[width=1.80\columnwidth]{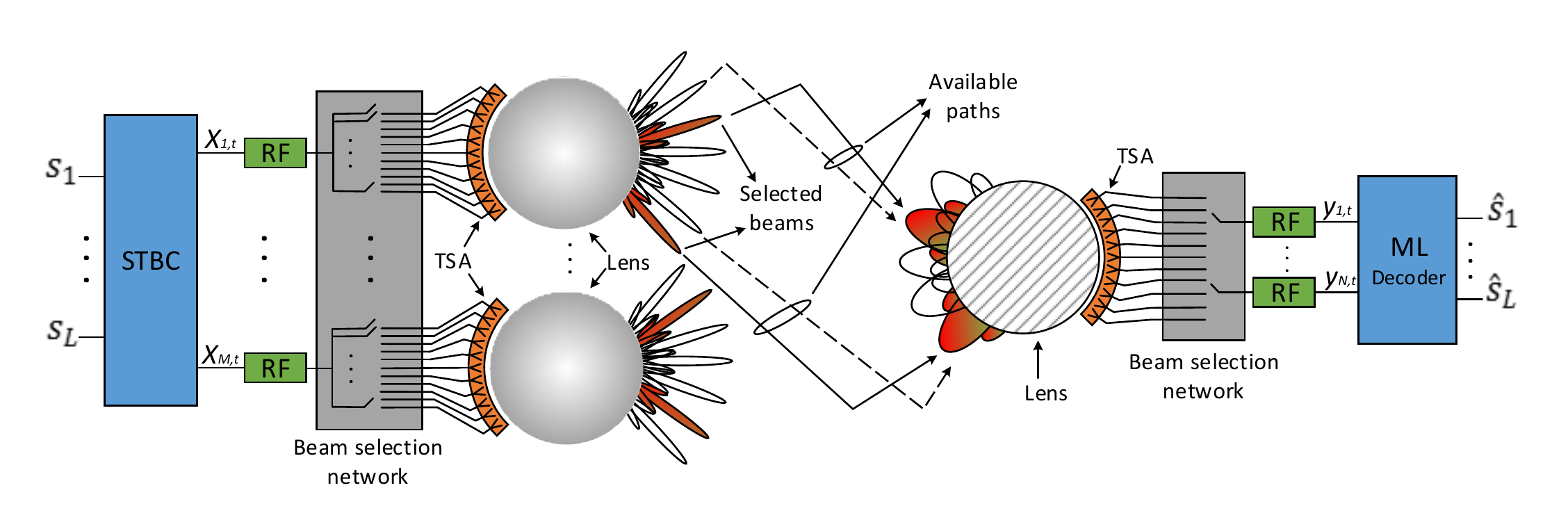}
    \vspace{-0.5cm}
    \caption{The proposed lens-based RA-MIMO system. Left hand side shows base station where complete lenses are used. Each lens is associated with a single RF chain. In right hand side an incomplete lens is used. Several RF chains are connected to the lens.}
    \label{fig2}
\end{figure*}

To keep the application of RA-MIMO systems feasible and compensate for using incomplete lens, we proposed a new RA-MIMO structure as shown in  Fig.~\ref{fig2} on the next page. At the transmitter side, there are multiple reconfigurable antennas. At each antenna, only one RF chain is connected to TSA arrays via BSN. In contrast,  at the receiver side multiple RF chains are allowed to be connected to the TSAs. As it is shown in Fig.~\ref{fig2},  the receiver can have multiple RF chains but only one incomplete lens. 

As mentioned,  each reconfigurable antenna contains multiple TSA arrays. In the reconfigurable antennas with complete lens, each TSA element is able to steer a narrow and strong beam~\cite{r19}.  On the other hand, Due to using incomplete lens, the beams steered by the receive antenna are wide. So, each wide beam at the receiver is able to capture several transmit narrow beams. This means that the number of channels between a transmit  reconfigurable antenna and a receive beam can be more than one (see Fig.~\ref{fig2}). 

Each RF chain at the transmitter selects several beams. While each RF chain at the receiver can only select one beam. The number of selected beams by each transmit RF chain is equal to the number of receive beams. More details are provided in Fig.~\ref{fig2} on the next page. Also,  parameters of the proposed MIMO system have been listed in Table~\ref{tab:my_label}. 

Notice that although there is one reconfigurable antenna at the receiver,  the proposed system still has MIMO structure. In fact,  since there are multiple RF chains at both transmitter and receiver,  the proposed system is able to simultaneously transmit/receive signals. 

\begin{table}
\caption{Parameters and description}
\begin{tabular}{ | m{1cm}| m{6.9cm} | } 
\hline
Parameter & Description \\ 
\hline
$M$ & Number of transmit reconfigurable antennas \\ 
\hline
$M_\text{TSA}$ & Number of TSA elements associated to each transmit antenna \\ 
\hline
$N_\text{TSA}$ & Number of TSA elements associated to each receive antenna \\ 
\hline
$M_\text{B}$ & Number of steering beams from each transmit antenna\\ 
\hline
$N_\text{B}$ & Number of steering beams from each receive antenna\\ 
\hline
$\mathcal{B}_{n, m}$ & Set of beams steered from $m$th transmit antenna to $n$th receive antenna\\
\hline
$B_{n, m}$ & Number of members in set $\mathcal{B}_{n, m}$\\
\hline
\end{tabular}
\label{tab:my_label}
\end{table}

Based on the structure of the proposed MIMO the following inequalities hold.
\begin{enumerate}
    \item $M_\text{B}\ll M_\text{TSA}$.
    \item $N_\text{B} \leq N_\text{TSA}$.
    \item $\sum_{n=1}^{N_\text{B}}\sum_{m=1}^M B_{n,m} \ll MM_\text{TSA}$.
\end{enumerate}
The first inequality follows from the fact that mmWave channels are sparse. Since the transmit antenna is equipped with complete lens, a huge number of TSAs can be located on the surface of the lens~\cite{r19}. So, the number of paths is far less then the number of TSAs. In the second inequality, due to using incomplete lens, the surface is small which means less TSAs can be located. Thus, the number of receive beams can be the same or less than the number of TSAs. The explanation for the first item is valid for the third one, too.       

Recall that each TSA at the transmitter side directs the beam toward only one of the beams at the receiver side. So, it gives 
\vspace{-10pt}
\begin{equation}
    \sum_{n=1}^{N_\text{B}}B_{n, m}=M_\text{B}.
\end{equation}
This assumption makes sense since the beams are highly directional and narrow. Thus,  each beam is captured by only one of beams at the receiver. Therefore,  we have 
\begin{equation}\label{eq:3}
    \mathcal{B}_{n, m} \cap \mathcal{B}_{n^\prime, m} = \emptyset, \quad \text{for} \quad n\neq n^\prime. 
\end{equation}

Denoting $h_{n, m, k}$ the channel coefficient corresponding to $k$th beam for $k=1, 2, \dots, B_{n,m}$,  the channel coefficient vector between the $m$th transmit antenna and $n$th receive beam is given by
\vspace{-10pt}
\begin{equation}\label{eq:4}
    \mathbf{h}_{n, m} = \left[h_{n, m, 1},  \dots,  h_{n, m, B_{n, m}}\right]^T.
\end{equation}
In this paper, the channel coefficients are assumed to experience Rayleigh fading with zero-mean and unit variance,  i.e.,  $\mathbb{E}\left[h_{n, m, k}\right]=0$ and $\mathbb{E}\left[\left|h_{n, m, k}\right|\right]=1$.  
Further, the MIMO channel matrix,  $\mathbf{H}$,  of size $N_\text{B}\times M$ is giving by
\begin{equation}\label{eq:5}
    \mathbf{H} = [\hbar_{n, m}], 
\end{equation}
where $\hbar_{n, m}$ is the selected channel between $B_{n, m}$ channels. 

\section{Beam Selection Technique}
Although the incomplete lens reduces the occupied room,  its performance correspondingly degrades~\cite{ebling2006multi}. Compared to a complete spherical lens,  the incomplete lens partially loses the amplification and directivity capabilities~\cite{ebling2006multi}. This issue leads to lower SNR and would make the application of RA-MIMO in mobile devices questionable. That is to say,  the primary aim of these MIMO systems,  which is to increase the reliability in mmWave frequencies,  might not be guaranteed. Hence, an appropriate beam selection algorithm is required to increase the SNR. Regarding the assumption that the BSN in each transmit antenna selects only one beam amongst all possible beams,  two cases are considered for beam selection,  random beam selection and optimal beam selection. It is assumed that in both cases the receiver selects intended beams and feeds them back to the transmitter.  

\subsection{RA-MIMO with Random Beam Selection}
The work in~\cite{almasi2018new} shows that combining STBCs with the RA-MIMO leads to rate-one and full-diversity gain. In the proposed RA-MIMO, to increase the order of diversity, the number of transmit antennas should be increased\footnote{Since we want to keep the number of RF chains at the receiver low.}. Unfortunately, this imposes high hardware complexity,  costs and power consumption. To partially overcome these issues, we constrain the number of transmit antennas to two, i.e., $M=2$. It is clear that the number of RF chains will be two.   

Recall that there are several paths from each transmit antenna to each of the beams at the receiver. At the receiver one of the paths is randomly selected and information of the corresponding angle of departure is fed back to the transmitter. Then the transmitter sends the information symbols $s_i$ for $i=1, 2$ to the receiver. The received $N_\text{B}\times 2$ matrix,  $\mathbf{Y}$,  is given by
\begin{equation}\label{eq:7}
\mathbf{Y} = \sqrt{\gamma_0}\mathbf{H}\mathbf{X} + \mathbf{N}, 
\end{equation}
where $\gamma_0$ is the average received SNR. $\mathbf{H}$ is defined in (\ref{eq:5}).  $\mathbf{N}$ of dimension $N_\text{B}\times 2$ denotes the noise matrix with entry $n_{n, m}$ which is independent identically distributed (i.i.d) random variable with $\mathcal{CN}(0, 1)$. The matrix $\mathbf{X}$ denotes Alamouti STBC given by~\cite{alamouti1998simple}\vspace{-10pt}
\begin{equation}\label{eq:8}
    \mathbf{X} = \begin{bmatrix}
    s_1 & s_2\\
    -s^*_2 & s^*_1
    \end{bmatrix}.
\end{equation}
It is easy to show that utilization of Alamuti STBC in the proposed MIMO gives the following average SNR.\vspace{-5pt}
\begin{equation}\label{eq:6}
    \mathbb{E}[\gamma_\text{ran}] = \gamma_0\sum_{n=1}^{N_\text{B}}\sum_{m=1}^{2}\mathbb{E}\left[\left|\hbar_{n, m}\right|^2\right]=2\gamma_0N_\text{B},
\end{equation}
where $\hbar_{n, m}$ is defined in (\ref{eq:5}).  The above equation states that only by increasing the number of RF chains at the receiver the SNR increases. 
\subsection{Optimal Beam Selection Algorithm}\label{sec:BS}
In traditional MIMO systems antenna selection combined with STBC has been exploited as a promising solution to increase the diversity order whereas keep the number of RF chains small~\cite{gore2002mimo,ma2007antenna}. Inspired by these works, here,  we study a beam selection technique for the proposed MIMO system. Of interest,  we evaluate the achievable gain of Alamouti STBC with optimal beam selection technique over the random beam selection. 

Let $z_{n, m}$ denote the maximum channel gain of vector $\mathbf{h}_{n, m}$ represented in~(\ref{eq:4}),  i.e.,     
\begin{equation}\label{eq:10}
    z_{n, m} = \underset{k}{\text{max}} \ \left|\mathbf{h}_{n, m}\right|^2,  \quad  \forall n, m. 
\end{equation}
That is,  the BSN selects the beam which gives the highest gain to maximize the received SNR. Since it is assumed that $M=2$,  we will have $2N_\text{B}$ maximum values of $z_{n,m}$. Then,  the received instantaneous SNR becomes
\begin{equation}\label{eq:16}
    \gamma_\text{opt} = \gamma_0\sum_{n=1}^{N_\text{B}}\sum_{m=1}^{2}z_{n, m}.
\end{equation}

The probability density function (PDF) of variable $z_{n, m}$ is calculated as~\cite{balakrishnan2014order}
\begin{equation}\label{eq:11}
    p_Z(z_{n, m}) = B_{n, m}(F_Z(z_{n, m}))^{(B_{n, m}-1)}f_Z(z_{n, m}), 
\end{equation}
where 
\begin{equation}\label{eq:12}
    f_Z(z_{n, m}) = e^{-z_{n, m}}, 
\end{equation}
is the density function of chi-square with two degrees of freedom. $F_Z(z_{n, m})$ denotes the cumulative density function (CDF) of chi-square distribution which is expressed as
\begin{equation}\label{eq:13}
   F_Z(z_{n, m}) = 1-e^{-z_{n, m}}. 
\end{equation}
 The expected value of variable $z_{n, m}$ is obtained as~\cite{gore2002mimo}
\begin{align}\label{eq:14}
    \mathbb{E}[z_{n, m}] = B_{n, m}\sum_{p=0}^{B_{n, m}-1}(-1)^p\left(\begin{smallmatrix}
    B_{n, m}-1\\
    p
  \end{smallmatrix}\right)\frac{1}{(p+1)^2}.
\end{align}
Following the selection gain definition in~\cite{gore2002mimo},  the achievable average SNR gain is given by
\begin{align}\label{eq:15}
    G \equiv 10\text{log}_{10}\left(\frac{\mathbb{E}[\gamma_\text{opt}]}{\mathbb{E}[\gamma_\text{ran}]}\right) &= 10\text{log}_{10}\left(\frac{\sum_{n=1}^{N_\text{B}}\sum_{m=1}^{2}\mathbb{E}[z_{n, m}]}{2N_\text{B}}\right)\nonumber \\
    &=10\text{log}_{10}\left(\mathbb{E}[z_{n, m}]\right),
\end{align}
where $\mathbb{E}[z_{n,m}]$ is given by~(\ref{eq:14}). Now, one can see that for $B_{n,m}>1$, the gain is always greater than one.   
\section{Numerical Results}\label{sec:simulation}
In this section,  we discuss the result of numerical simulations for the proposed RA-MIMO system in Fig.~\ref{fig2} and beam selection algorithm in~(\ref{eq:10}). The entries of channel matrix in~(\ref{eq:5}) is modeled as Rayleigh fading channel with $\mathcal{CN}$(0, 1). 

\begin{figure}
\vspace{-0.2cm}
    \centering
    \includegraphics[scale=0.5]{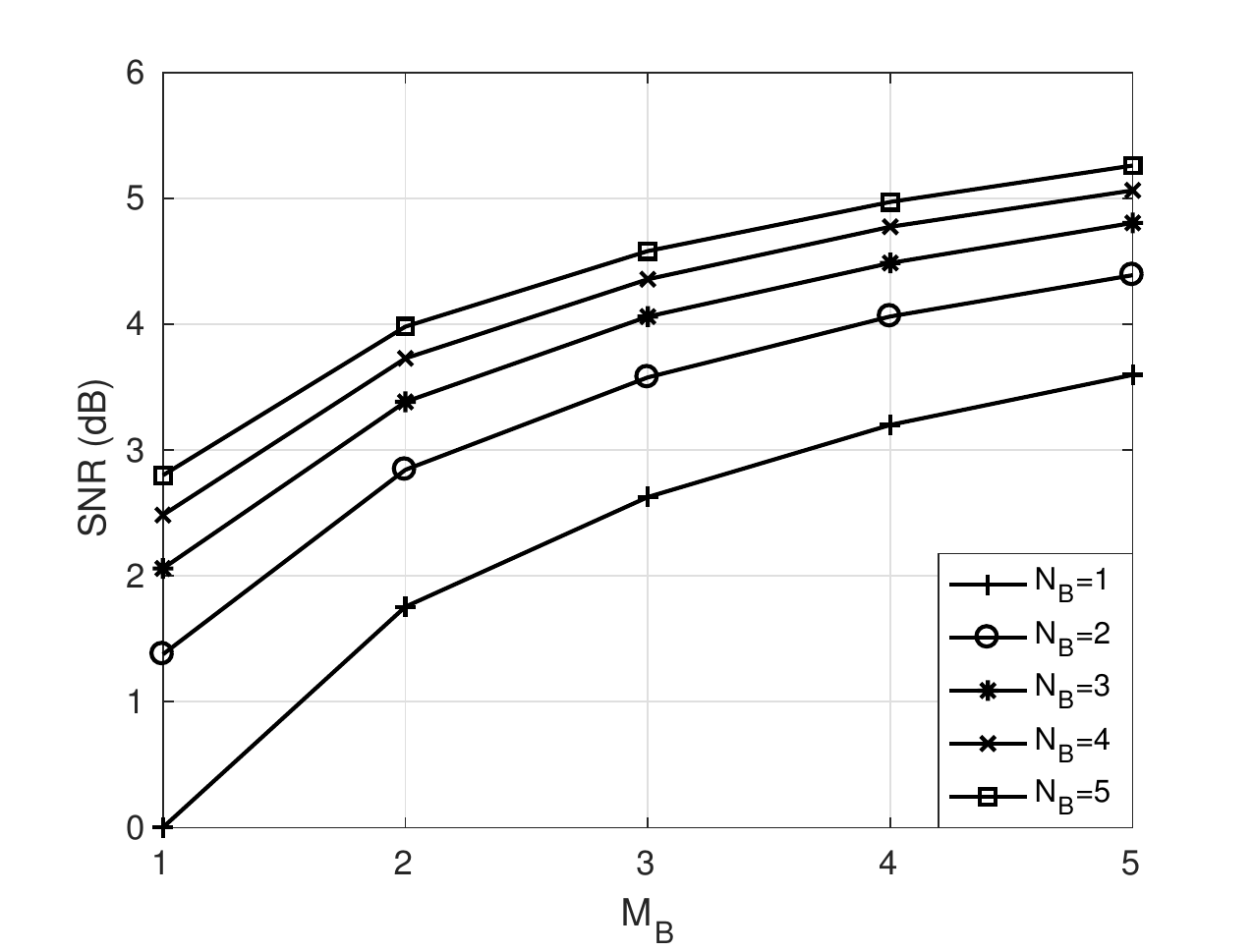}
    \vspace{-0.3cm}
    \caption{The average SNR gain versus number of steering beams from each transmit antennas for various number of beams at the receiver.}
    \label{fig:SNR}
\end{figure}

Figure~\ref{fig:SNR} shows the attainable average SNR gain for various number of beams at the transmitter and the receiver. According to the beam selection algorithm, the beams with higher gains are selected. Therefore, for a fixed number of steering beams at the receiver,  i.e.,  $N_\text{B}$,  by increasing the number  of steering beams at the transmitter,  i.e.,  $M_\text{B}$,  the gain becomes larger. This follows since the beam selection network chooses beams with the highest gain. Obviously,  when the number of steering beams increases,  the probability that a channel with high gain occurs becomes high. Similarly,  for large $N_\text{B}$,  the gain increases.  

\begin{figure}
\vspace{-0.2cm}
    \centering
    \includegraphics[scale=0.5]{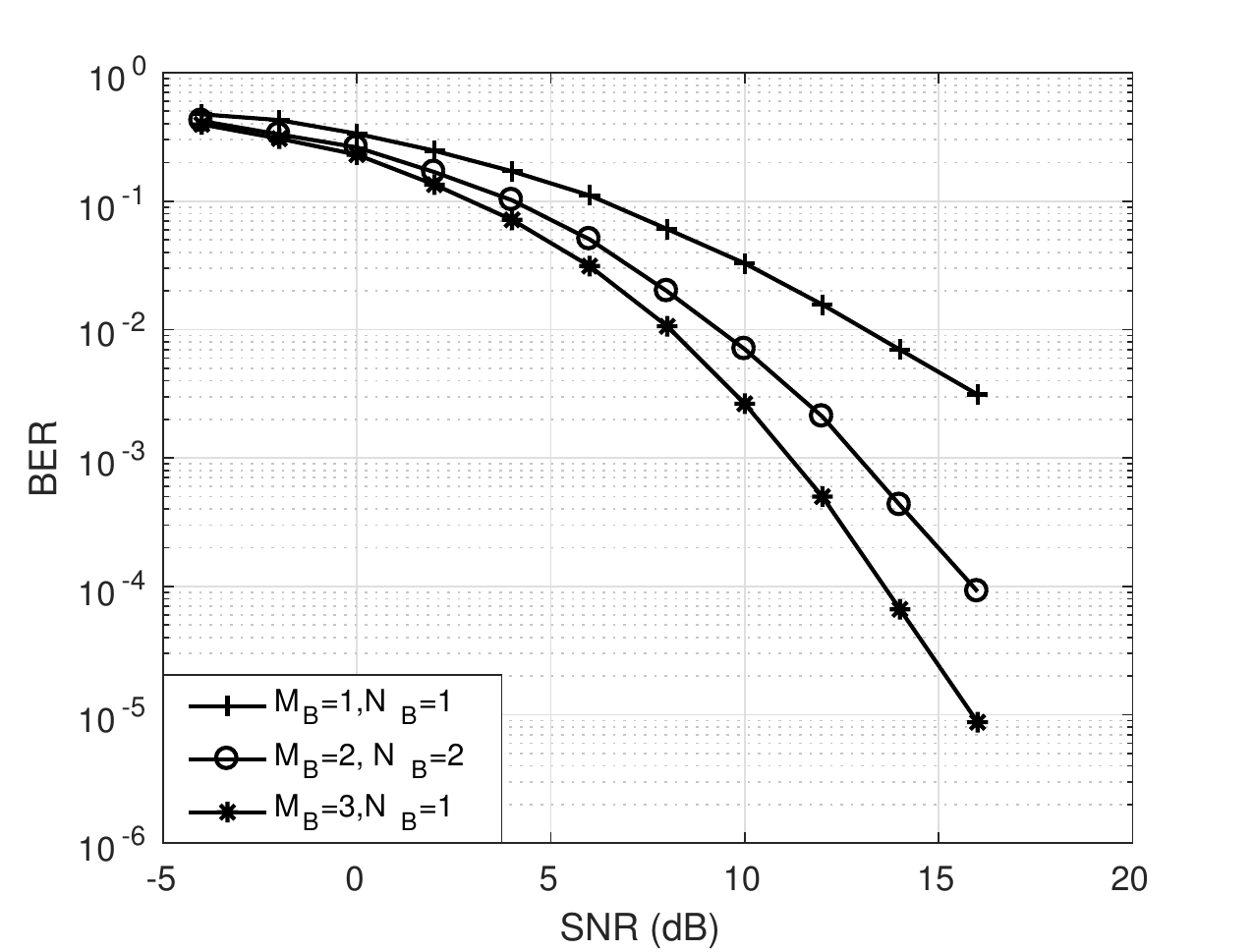}
    \vspace{-0.3cm}
    \caption{BER performance versus SNR for $2\times1$ RA-MIMO integrated with the optimal beam selection algorithm.}
    \label{fig:BER}
\end{figure}

Figure~\ref{fig:BER} compares the BER performance of random beam selection and the optimal beam selection algorithm. Considering the transmitter is equipped with two reconfigurable antennas and the receiver orients only one beam,  the proposed RA-MIMO along with Alamouti STBC and optimal beam selection algorithm achieves full-diversity gain. For instance,  for $M_\text{B}=1$ and $N_\text{B}=1$,  the diversity is of order 2 as it is expected. Also,  for $M_\text{B}=$ 2 and 3, the system attains the diversity of order 4 and 6,  respectively. This means that to achieve high diversity gain,  the system does not need to increase size of the MIMO.


\section{Conclusion}\label{sec:conclusion}
In this paper,  we proposed a MIMO system by using  lens-based multi-beam reconfigurable antennas. To suppress required large space at user device, we used a incomplete lens. The proposed system is able to establish a large number of paths between the transmitter and the receiver which overcomes the channel sparsity. Further,  to suppress high path loss and hardware complexity,  Alamouti STBC is integrated with the proposed optimal beam selection. The analysis and simulations verify the efficiency of the proposed RA-MIMO system in mmWave band wireless communications.

\appendices




\ifCLASSOPTIONcaptionsoff
  \newpage
\fi



%
\bibliographystyle{IEEEtran}
\bibliography{IEEEabrv,main}

%







\end{document}